\documentclass[journal=apchd5,manuscript=article]{achemso}

\usepackage[version=3]{mhchem} 
\usepackage{amsmath}
\usepackage{xcolor}
\usepackage[T1]{fontenc}  
\usepackage{textcomp}     
                  
\title{Boosting Fluorescence and Polarization\\ with Nanophotonics for\\ Single-Molecule Chiral Sensing}

\author{Ershad Mohammadi}
\affiliation{
Department of Applied Physics and Eindhoven Hendrik Casimir Institute,
Eindhoven University of Technology, Eindhoven, The Netherlands
}
\alsoaffiliation{
Van der Waals--Zeeman Institute, Institute of Physics,
University of Amsterdam, Amsterdam, The Netherlands
}

\author{T. V. Raziman}
\affiliation{
Department of Applied Physics and Eindhoven Hendrik Casimir Institute,
Eindhoven University of Technology, Eindhoven, The Netherlands
}
\alsoaffiliation{
Blackett Laboratory, Department of Physics,
Imperial College London, London, UK
}
\alsoaffiliation{
Department of Mathematics,
Imperial College London, London, UK
}

\author{Alberto G. Curto}
\affiliation{
Photonics Research Group, Ghent University--imec,
Ghent, Belgium
}
\alsoaffiliation{
Center for Nano- and Biophotonics,
Ghent University, Ghent, Belgium
}
\alsoaffiliation{
Department of Applied Physics and Eindhoven Hendrik Casimir Institute,
Eindhoven University of Technology, Eindhoven, The Netherlands
}
\email{Alberto.Curto@UGent.be}


\begin{document}

\maketitle

\begin{abstract}
Detecting chirality is essential in chemistry and biomedicine. Circular dichroism identifies chiral molecules through their differential absorption of circularly polarized light. However, its sensitivity is limited by vanishing absorption at low concentrations.
Fluorescence-based methods have already enabled other optical techniques to reach the single-molecule limit and could, in principle, also extend chiral detection to reach this sensitivity regime. They are marred, however, by the combination of weak fluorescence signals and low polarization contrast.
To overcome these limitations, we leverage high-index dielectric nanophotonics to simultaneously enhance both fluorescence and polarization for chiral sensing.
We propose a reciprocity-based analytical framework for a chiral dipolar emitter coupled to a nanoantenna and reveal the different contributions and trade-offs governing the enhancement of circularly polarized fluorescence.
Guided by this insight, we tailor achiral dielectric nanostructures to reach a 1000-fold enhancement of fluorescence intensity together with a 10-fold increase in polarization.
Crucially, we show that accounting for extrinsic polarization effects in the angular emission patterns of nanophotonic structures is indispensable for reliable enantiomer discrimination.
Our approach opens a realistic path for single-molecule chiral sensing.
\end{abstract}

\section{Chiral sensing at low molecular concentrations}

Chirality plays a key role in biochemistry, underpinning a wide range of processes in the pharmaceutical and food industries~\cite{francotte2006chirality,zhou2018chiral,kurihara1998chirality}. Enantiomers are molecules that exist as non-superimposable mirror images, and their handedness strongly influences their biological function. Therefore, distinguishing between enantiomers is crucial in areas such as drug development and quality control~\cite{crossley1995chirality}. 
Common chiroptical spectroscopies, such as optical rotation and circular dichroism (CD), are particularly accurate when large concentrations of chiral molecules are available~\cite{schellman1975circular}. Nanophotonics offers a powerful route to increase the sensitivity of absorption-based chiral sensing by generating superchiral near fields~\cite{schaferling2017chiral,solomon2020nanophotonic, im2024perspectives,yoo2019metamaterials,mohammadi2019accessible,mohammadi2018nanophotonic, raziman2019conditions, mohammadi2021dual} or through chirality transfer mechanisms~\cite{mohammadi2023nanophotonic}. However, pushing the detection limit down to the single-molecule level remains extremely challenging. Such an ultrasensitive chiroptical analysis could uncover the conformational dynamics of individual biomolecules, usually masked by ensemble averaging.

The use of fluorescence for chiral sensing provides background-free molecular detection thanks to the spectral separation between excitation and emission wavelengths. It is the method of choice for detecting single achiral molecules and thus offers a route toward single chiral molecule imaging and spectroscopy. 
The main fluorescence-based chiroptical techniques are fluorescence-detected circular dichroism (FDCD) and circularly polarized luminescence (CPL)~\cite{tinoco1977fluorescence, turner1974fluorescence,tang2009limits, gaiduk2010room}. Both FDCD and CPL combine the chiral specificity of absorption-based CD with the background-free sensitivity of fluorescence. FDCD measures absorption CD indirectly through differences in fluorescence under right- and left-handed circularly polarized excitation. On the other hand, CPL exploits the dissymmetry in circularly polarized emission of a chiral molecule.  However, both techniques ultimately suffer from weak chiral light–matter interactions. Achieving a high signal-to-noise ratio requires not only strong polarization contrast but also sufficiently intense fluorescence, motivating the need for their simultaneous enhancement leading to higher chiral brightness~\cite{arrico2021quantifying}. 

Nanophotonic structures, made of metals or high-index dielectrics, can enhance fluorescence through localized resonances~\cite{anger2006fluor, curto2010unidirectional,kuhn2006enhancement,tam2007plasmonic,cihan2018silicon,murai2020enhanced,bidault2019dielectric,feng2017isotropic} by increasing the excitation rate, quantum efficiency, and collection efficiency~\cite{sun2016fluorescence}.
While fluorescence enhancement has been studied for electric and magnetic dipole emitters separately~\cite{anger2006fluor,kuhn2006enhancement,feng2017isotropic}, emission-based chiral sensing involves a combination of dephased electric and magnetic dipoles (i.e., a chiral dipole). High-index dielectric nanostructures are particularly promising to achieve the required polarization control because they can be tailored to preserve the helicity of light in their near field \cite{graf2019achiral} through their electric and magnetic Mie resonances~\cite{zambrana2016tailoring}. There is a trade-off, however, between the enhancement of fluorescence intensity and polarization contrast. In the seminal demonstration by Tang and Cohen, the chiroptical contrast was enhanced using superchiral fields produced near a node of the electric field~\cite{tang2011enhanced}. The suppression of the electric-field intensity inherently reduced molecular excitation and, consequently, the fluorescence signal. This suppressed emission has been a key limitation towards achieving sensitivity at the single chiral molecule level. More recently, Solomon \textit{et al.} employed a silicon metasurface to enhance the FDCD response of chiral molecular monolayers through increased local optical chirality~\cite{solomon2020fluorescence}. 

Here, focusing on circularly polarized luminescence, we develop dielectric nanoresonators to simultaneously boost fluorescence and polarization from single chiral molecules. In CPL, chirality is encoded in the polarization state of the emitted light and quantified by the degree of circular polarization (DOCP). First, we derive a rigorous analytical framework based on the reciprocity theorem to describe the interaction between a chiral emitter and a general nanoantenna. From this analysis, we formulate closed-form expressions that relate the degree of circular polarization to key near-field quantities like optical chirality and electric energy density. These insights enable the rational design of nanophotonic structures tailored for fluorescence-based chiral sensing, leading to achiral high-index dielectric nanostructures that enhance circularly polarized fluorescence. Our analysis further reveals that the extrinsic polarization induced by the nanostructure must be carefully accounted for to distinguish the intrinsic polarization of opposite enantiomers. Finally, a multipolar expansion explains the underlying contributions to intensity and polarization enhancements. By establishing the theoretical foundations for brighter CPL, our results enable a powerful approach to fluorescence-based chiral sensing toward single-molecule chiroptical detection. 

\begin{figure}[t!]
    \centering
    \includegraphics[width=\textwidth]{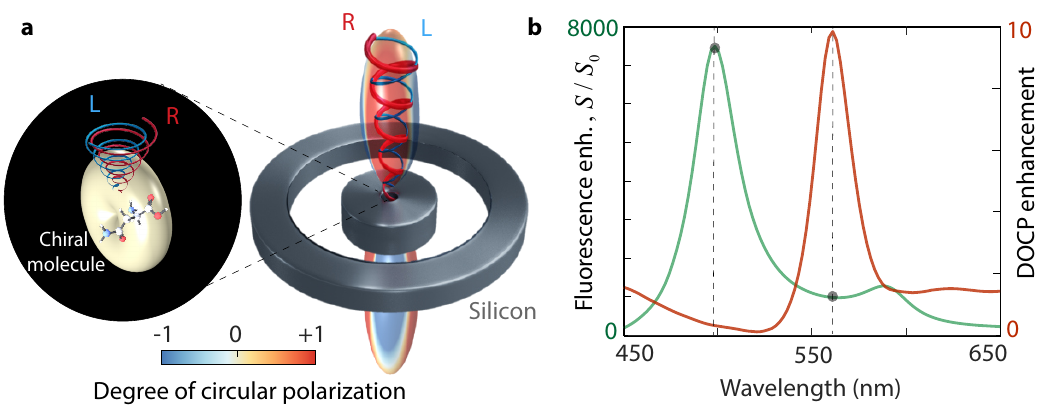}
    \caption{\textbf{Boosting circularly polarized fluorescence from a chiral molecule with a high-index dielectric nanostructure}. 
(a) A fluorescent chiral molecule with a small dissymmetry factor emits nearly equal right- and left-handed circularly polarized light in free space (R and L; red and blue helices). A holey disk–ring nanoresonator enhances both the fluorescence intensity and the polarization contrast of the emitter. 
(b) Enhancements of fluorescence and degree of circular polarization (DOCP) for a chiral emitter with dissymmetry parameter $\alpha = 0.005$ and intrinsic quantum yield $\eta = 0.001$. The structure provides approximately one order-of-magnitude enhancement in DOCP at $\lambda = 560$~nm, together with a 1000-fold enhancement in fluorescence at the same wavelength. Alternatively, it can provide a 7500-fold enhancement in fluorescence at an excitation wavelength of 495~nm, while decreasing the DOCP compared to the initial case.}
  \label{concept}
\end{figure}


\section{Enhanced fluorescence and polarization}

Circularly polarized fluorescence is characterized by the fluorescence intensity ($S$) and the degree of circular polarization (DOCP), which ranges from $-1$ to $1$. For a chiral molecule coupled to a nanostructure, the right-($+$) and left-($-$)handed circularly polarized components of the fluorescence are
\begin{equation}\label{SF}
S_{\pm} \propto \xi \, A_{\mathrm{ex}} \, \eta_{\pm},
\end{equation}
where $\xi$ is the collection efficiency, defined as the fraction of the emitted light collected by the detection system~\cite{mohammadi2008gold}, $A_{\mathrm{ex}}$ is the excitation rate, and $\eta_{\pm}$ are the quantum yields associated with right- and left-handed emission channels. Throughout this work, we adopt the time-harmonic convention $e^{i\omega t}$, where $\omega$ is the angular frequency. 

The excitation rate is given by $A_{\mathrm{ex}} = \omega \mathrm{Im} \left( \mathbf{E}_{\mathrm{ex}} \cdot \mathbf{p}_{\mathrm{ex}}^{*} + \mathbf{B}_{\mathrm{ex}} \cdot \mathbf{m}_{\mathrm{ex}}^{*} \right)/2$, where $\mathbf{p}_{\mathrm{ex}}$ ($\mathbf{m}_{\mathrm{ex}}$) denotes the electric (magnetic) dipole moment at the excitation wavelength $\lambda_{\mathrm{ex}}$, and $\mathbf{E}_{\mathrm{ex}}$ ($\mathbf{B}_{\mathrm{ex}}$) represents the corresponding local field at the dipole position.~\cite{tang2010optical} Neglecting the magnetic polarizability relative to the electric polarizability reduces the excitation rate to $A_{\mathrm{ex}} = \omega \mathrm{Im}(\alpha_e) |\mathbf{E}_{\mathrm{ex}}|^2/2$, where $\alpha_e$ is the electric polarizability. Consequently, the excitation rate is directly proportional to the intensity of the electric field at the excitation wavelength.

The quantum yields are given by $\eta_{\pm} = \gamma_{\pm}^{\mathrm{rad}} / (\gamma_{+}^{\mathrm{rad}} + \gamma_{-}^{\mathrm{rad}} + \gamma^{\mathrm{nrad}})$, where $\gamma_{\pm}^{\mathrm{rad}}$ are the radiative decay rates for right- and left-handed emission, and $\gamma^{\mathrm{nrad}}$ is the nonradiative decay rate. The total fluorescence intensity is $S = S_{+} + S_{-}$, corresponding to a total quantum efficiency of $\eta = \eta_{+} + \eta_{-}$ and a total radiative decay rate of $\gamma = \gamma_{+}^{\mathrm{rad}} + \gamma_{-}^{\mathrm{rad}}$.

The degree of circular polarization is defined as 
\begin{equation}\label{DOCP}
\mathrm{DOCP} = \frac{S_{+} - S_{-}}{S_{+} + S_{-}} = \frac{\gamma_{+}^{\mathrm{rad}} - \gamma_{-}^{\mathrm{rad}}}{\gamma_{+}^{\mathrm{rad}} + \gamma_{-}^{\mathrm{rad}}}.
\end{equation}
For chiral molecules in the absence of a nanostructure, the same expressions remain valid with the corresponding quantities denoted by the subscript $0$. Due to the inherently weak chiral light--matter interaction, the radiative decay rates $\gamma_{0,+}^{\mathrm{rad}}$ and $\gamma_{0,-}^{\mathrm{rad}}$ for typical organic molecules differ only slightly. Consequently, the intrinsic DOCP is very small, typically on the order of $10^{-4}$--$10^{-2}$ (Figure~\ref{concept}a, inset).

An optimal nanophotonic structure for boosting circularly polarized fluorescence, according to Equations~\ref{SF} and ~\ref{DOCP}, should maximize the collection efficiency by increasing the emission directivity beyond the characteristic doughnut-shaped radiation pattern of a chiral dipolar emitter in free space. It should also strongly increase the excitation rate to maximize fluorescence. Finally, it should improve the quantum yields of both polarization channels, $\eta_{+}$ and $\eta_{-}$, while maintaining a large asymmetry between them or even increasing it to achieve a larger DOCP.

To reach these goals, we propose a silicon holey disk surrounded by a larger concentric ring (Figure~\ref{concept}a). The disk and ring concentrate the magnetic field at the central symmetry point through constructive interference of radiative magnetic modes~\cite{yang2017boosting,darvishzadeh2019optical,li2017enhancing,yang2020strong}. As we demonstrate below, this enhanced magnetic field increases the polarization contrast. Simultaneously, the same modes also boost the emission directivity, enabling efficient fluorescence collection. In Figure~\ref{concept}, the holey disk was optimized to have a radius and height of 80 and 90~nm, respectively, while the central hole has a radius of 10~nm. The surrounding ring has inner and outer radii of 320 and 420~nm and the same height as the disk. We perform full-wave finite-element simulations in COMSOL Multiphysics 5.5 using realistic silicon permittivity values~\cite{aspnes1983dielectric}. 
For simplicity, we model the nanostructure as an isolated component suspended in air. In practice, nanostructures could also be arranged on a glass substrate in periodic arrays without qualitatively altering the main observations. Guided by the analytical considerations in Supporting Section~S2, we identify the center of the hole as the optimal position for the chiral molecule.
Importantly, the symmetry of this location ensures that reversing the molecular handedness reverses the CPL signal for radiation normal to the structure, indicating the absence of extrinsic polarization contributions.

The emission of a fluorescent chiral molecule can be modeled as a superposition of parallel electric and magnetic dipoles with a relative phase of $\pi/2$, written as $\sigma\hat{n} = (\mathbf{p} \pm i \alpha \mathbf{m})\hat{n}$, where $\hat{n}$ is the unit vector along the dipole direction and the $\pm$ sign corresponds to right- and left-handed enantiomers~\cite{eismann2018exciting}.  To ensure equal radiated power of the electric ($\mathbf{p}$) and magnetic ($\mathbf{m}$) dipoles, we impose the condition $\mathbf{m} = c_{0}\mathbf{p}$, where $c_{0}$ is the speed of light in free space. The relative strength of the magnetic contribution is then controlled by the dissymmetry parameter $\alpha$. For a chiral dipolar emitter in a homogeneous medium, the DOCP is angularly uniform and equals $2\alpha/(1+\alpha^{2})$ (Supporting Section~S1). Furthermore, the right- and left-handed circularly polarized components of the emitted power are proportional to the radiative decay rates of a chiral molecule, $\gamma_{0,+}^{\mathrm{rad}}$ and $\gamma_{0,-}^{\mathrm{rad}}$, respectively. We quantify the enhancements provided by the nanostructure relative to the homogeneous-medium case by calculating the fluorescence and DOCP enhancements for an $x$-oriented chiral dipole coupled to the disk--ring system. 

Our proposed disk--ring nanostructure simultaneously enhances the circular polarization and fluorescence of a weakly chiral emitter, yielding a 10-fold increase in DOCP and a 1000-fold increase in fluorescence intensity at $\lambda_{\mathrm{em}} = 560$ nm (Figure~\ref{concept}b). The molecule is excited at $\lambda_{\mathrm{ex}} = 495$~nm and has an intrinsic dissymmetry parameter $\alpha = 0.005$, corresponding to a free-space DOCP of 0.01. The DOCP upon coupling to the nanostructure thus becomes 0.1. We obtain this value by considering a numerical aperture of $\mathrm{NA} = 0.2$ and integrating the right- and left-handed circularly polarized components of the radiated power over the collection cone ($\theta \leq \theta_{\mathrm{NA}} = \sin^{-1}(\mathrm{NA})$). These powers are proportional to the corresponding radiative decay rates of the chiral emitter, $\gamma_{\mathrm{rad},+}$ and $\gamma_{\mathrm{rad},-}$, respectively (see Equation~\ref{DOCP}). We analyze the dependence of these results on dipole orientation in Supporting Section~S3. 

\begin{figure}[t!]
    \centering
    \includegraphics[width=\textwidth]{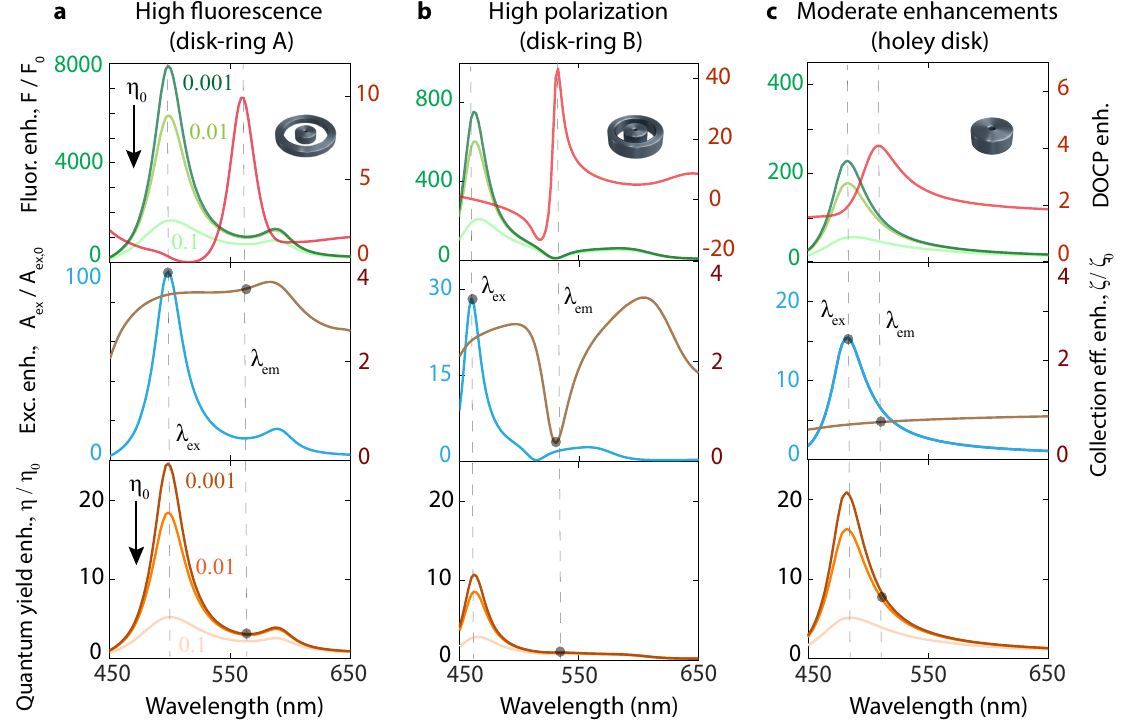}
    \caption{\textbf{Contributions to circularly polarized fluorescence for different nanostructures}. Top: fluorescence and DOCP enhancements (green and red). Center: excitation and collection efficiency enhancements (blue and brown). Bottom: quantum efficiency enhancement (orange) for an $x$-oriented chiral emitter placed at the center of the hole in each structure, with parameters defined in the text. The green and orange curves represent increasing values of the intrinsic molecular quantum yield for a fixed intrinsic DOCP = 0.01 ($\alpha = 0.005$).}
     \label{Sf_Exci_Ce_Qe}
\end{figure}

To quantify the contributions to the fluorescence enhancement, we decompose it in Figure~\ref{Sf_Exci_Ce_Qe} into excitation ($A_{\mathrm{ex}} / A_{\mathrm{ex},0}$), collection efficiency ($\xi / \xi_{0}$), and quantum efficiency ($\eta / \eta_{0}$) according to Equation~\ref{SF}. We first focus on the disk--ring structure in Figure~\ref{concept} (disk--ring A) and plot the fluorescence and DOCP enhancements (green and red). Excitation dominates the fluorescence enhancement, followed by quantum efficiency, while collection efficiency provides a smaller contribution.
We obtain the excitation enhancement by illuminating the nanostructure with a linearly polarized plane wave at $\lambda_{\mathrm{ex}} = 500$~nm propagating along the $z$-direction. It reaches a factor of 90 (blue) and it is the main contributor to the overall fluorescence enhancement at $\lambda_{\mathrm{em}} = 560\,\mathrm{nm}$.
The collection efficiency in the normal direction for an $x$-oriented chiral dipole in free space is $\xi_{0} = 0.015$ (Supporting Section~S1). With the nanostructure, the directional radiation pattern leads to a 3.4-fold enhancement in collection efficiency (brown). 
The quantum efficiency enhancement is given by $ \left[ (1-\eta_{0})/F + \eta_{0}/\eta_{a} \right]^{-1}$, where $F$ and $\eta_{a}$ denote the Purcell factor ($F = \gamma^{\mathrm{rad}} / \gamma_{0}^{\mathrm{rad}}$) and the antenna efficiency~\cite{mohammadi2008gold,wientjes2014enhance}. The combination of a large Purcell factor and high antenna efficiency results in a 3.5-fold increase for an intrinsic quantum yield $\eta_{0} = 0.001$ (orange). As $\eta_{0}$ increases, the Purcell enhancement decreases and the quantum efficiency of the system gradually approaches the antenna efficiency $\eta_{a}$, reducing the overall fluorescence enhancement.

Higher polarization enhancements are also possible, but at the expense of reduced fluorescence enhancement, as explained through analytical derivations of the DOCP in the next
section. To exemplify this trade-off, we consider a second design with a higher aspect ratio (disk--ring B; Figures~\ref{Sf_Exci_Ce_Qe}b). In this configuration, the inner disk has a radius of 50~nm and a height of 150~nm with a central hole of radius 10 nm, while the ring has inner and outer radii of 250 and 350~nm. This structure is optimized for extreme polarization, yielding a remarkable 40-fold enhancement in DOCP at $\lambda_{\mathrm{em}} = 530$~nm (red). However, the reduced electric field (blue) significantly degrades both the collection efficiency (brown) and the quantum efficiency at that wavelength (orange), ultimately leading to a fluorescence enhancement of only 38 times (green).

Finally, one may opt for designs that provide more balanced enhancements in both fluorescence and polarization. For instance, a holey disk with a radius of 80~nm, a height of 70~nm, and a central hole of radius 10~nm yields a fourfold increase in polarization alongside a 250-fold enhancement in fluorescence (Figure~\ref{Sf_Exci_Ce_Qe}c). 
This structure provides a 15-fold enhancement in the excitation rate at $\lambda_{\mathrm{ex}} = 490$~nm (blue), while the collection efficiency is reduced by a factor of 0.75 at $\lambda_{\mathrm{em}} = 510$~nm (brown). The corresponding quantum efficiency enhancement reaches 7.5 (orange), which is higher than that of disk--ring A and B. However, this improvement remains insufficient to compensate for the lower excitation and collection efficiency enhancements provided by the disk--ring design.

\section{Reciprocity analysis of polarization enhancement}

To identify the mechanisms underlying polarization enhancement, we use the reciprocity theorem~\cite{balanis1999advanced} to model the interaction between a chiral emitter and a general nanoantenna. It relates the far-field response of the combined system---consisting of the chiral dipole and the nanophotonic structure---to the near field of the nanostructure alone (in the absence of the dipole) under appropriate illumination conditions. This relation provides a direct connection between the local electromagnetic field and the emitted polarization. Using this framework, we derive the DOCP emitted in the normal direction by an $x$-oriented, right-handed chiral emitter  (Supporting Section S2):
\begin{equation}\label{DOCP_General}
\mathrm{DOCP}(\sigma\hat{x}) =
\frac{\left|\mathrm{E}_{x,\mathrm{R}} + i\alpha Z_{0}\mathrm{H}_{x,\mathrm{R}}\right|^2 - \left|\mathrm{E}_{x,\mathrm{L}} + i\alpha Z_{0}\mathrm{H}_{x,\mathrm{L}}\right|^2}
{\left|\mathrm{E}_{x,\mathrm{R}} + i\alpha Z_{0}\mathrm{H}_{x,\mathrm{R}}\right|^2 + \left|\mathrm{E}_{x,\mathrm{L}} + i\alpha Z_{0}\mathrm{H}_{x,\mathrm{L}}\right|^2},
\end{equation}
where $Z_{0}$ is the free-space impedance, and $\mathrm{E}_{x}$ ($\mathrm{H}_{x}$) denotes the $x$-component of the electric (magnetic) near field at the dipole location when the nanostructure is illuminated by a circularly polarized plane wave. The subscripts $\mathrm{R}$ and $\mathrm{L}$ indicate the near fields generated under circularly polarized illumination with incident fields $\mathbf{E}_{\mathrm{R}/\mathrm{L}} = E_{\mathrm{inc}} \left( \hat{\mathbf{x}} \mp i \hat{\mathbf{y}} \right) \exp(i k_0 z)$.

The reciprocity treatment becomes particularly insightful when the emitter is positioned at a symmetry point of an achiral nanostructure, where the physical origin of polarization enhancement becomes clearer. Equation~\ref{DOCP_General} applies to both chiral and achiral nanophotonic structures, as well as to arbitrary dipole positions. However, at a symmetry point of an achiral nanostructure (e.g., the center of the hole in the disk and disk--ring configurations), symmetry imposes the conditions $\mathrm{E}_{x,\mathrm{R}}=\mathrm{E}_{x,\mathrm{L}}$ and $\mathrm{H}_{x,\mathrm{R}}=-\mathrm{H}_{x,\mathrm{L}}$. Under these conditions, Equation~\ref{DOCP_General} simplifies to:
\begin{equation}\label{DOCP_SymPoint}
\mathrm{DOCP}(\sigma\hat{x}) =
\frac{2\alpha Z_{0}\,\mathrm{Im}\!\left(\mathrm{E}_{x,\mathrm{R}}\mathrm{H}_{x,\mathrm{R}}^{*}\right)}
{|\mathrm{E}_{x,\mathrm{R}}|^2 + |\alpha Z_{0}\mathrm{H}_{x,\mathrm{R}}|^2},
\end{equation}
which gives the polarization emitted in the normal direction by a chiral emitter located at the symmetry point. Assuming a small dissymmetry parameter $\alpha$, the magnetic contribution in the denominator can be neglected ($|\mathrm{E}_{x,\mathrm{R}}|^2 \gg |\alpha Z_{0}\mathrm{H}_{x,\mathrm{R}}|^2$). In addition, the intrinsic DOCP of the chiral emitter in free space is approximately $2\alpha$. In this limit, the DOCP enhancement becomes:
\begin{equation}\label{DOCP_SymPoint_Cenh_E_enh}
\mathrm{DOCP}_{\mathrm{enh}}(\sigma\hat{x}) =
\frac{{C}/{C}_{\mathrm{inc}}}{\mathrm{E}^{2}/\mathrm{E}_{\mathrm{inc}}^{2}},
\end{equation}
where $\mathrm{E}$ and ${C}$ denote the electric field amplitude and the optical chirality of the near field at the dipole location under right-handed circularly polarized illumination, respectively. The optical chirality of the incident field is given by ${C}_{\mathrm{inc}} = k_{0}\mathrm{E}_{\mathrm{inc}}^{2}/(2c_{0}Z_{0})$, with $k_{0}$ being the free-space wavenumber.

The DOCP enhancement scales thus directly with optical chirality and inversely with electric energy density. This relation parallels the enhancement of the absorption dissymmetry factor, for which superchiral fields can increase enantiomeric contrast beyond that produced by circularly polarized plane waves~\cite{tang2010optical}. Here, the same ratio governs the emitted polarization, leading to superchiral emission with a higher DOCP than in a homogeneous medium. 

Two design strategies follow from this relation. First, reducing the electric field at $\lambda_{\mathrm{em}}$ increases the DOCP, as illustrated in Figure~\ref{Sf_Exci_Ce_Qe}b, but also suppresses fluorescence, which is undesirable for ultrasensitive detection. Second, the optical chirality at $\lambda_{\mathrm{em}}$ can be enhanced. However, increasing it through the electric field alone is ineffective because the electric energy density in the denominator of Equation~\ref{DOCP_SymPoint_Cenh_E_enh} grows more rapidly, reducing the DOCP. Magnetic-field enhancement, instead, increases the optical chirality without directly increasing the electric energy density, thereby enabling simultaneous enhancement of fluorescence intensity and polarization. An optimal nanophotonic structure should therefore enhance the electric field at both $\lambda_{\mathrm{ex}}$ and $\lambda_{\mathrm{em}}$ to increase fluorescence, while providing sufficient magnetic-field enhancement at $\lambda_{\mathrm{em}}$ to maximize the optical chirality relative to the electric energy density.

\begin{figure}[t!]
    \centering
        \includegraphics[width=\textwidth]{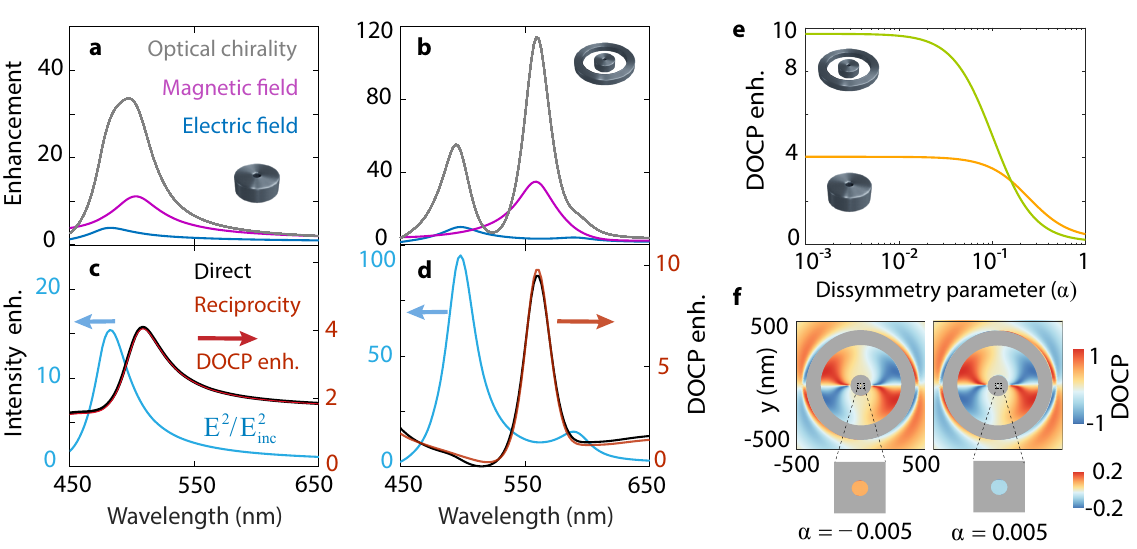}
    \caption{\textbf{
    Reciprocity framework for enhanced chiral emission.}
(a,b) Electric, magnetic, and optical chirality enhancements (blue, magenta, grey) for the holey disk and disk--ring A configurations, evaluated at the symmetry point under normal illumination by a circularly polarized plane wave. 
(c,d) Comparison of the polarization enhancements calculated through reciprocity using Equation~\ref{DOCP_SymPoint_Cenh_E_enh} (black) and direct numerical simulations (red), displaying the DOCP increase in the normal direction for a chiral emitter at the symmetry point. The electric field intensity enhancement (blue) is shown for reference.
(e) DOCP enhancement as a function of intrinsic dissymmetry parameter $\alpha$. 
(f) DOCP for different chiral emitter positions in the disk--ring A configuration computed using Equation~\ref{DOCP_General} at mid-height of the structure. The opposite enantiomers shown demonstrate a dominant intrinsic polarization inside the hole and extrinsic effects at other positions.}
    \label{reciprocity}
\end{figure}

High-index dielectric nanostructures supporting Mie-type electric and magnetic resonances with a dominant magnetic response are particularly well suited to realize these conditions.
We investigate this by calculating the electric, magnetic, and optical chirality enhancements for the holey disk and the disk--ring A configurations (Figure~\ref{reciprocity}a and b, respectively). Both structures exhibit strong optical chirality, primarily driven by pronounced magnetic field enhancements at 510~nm for the disk and 560~nm for the disk--ring A. At these wavelengths, the electric field enhancement remains sufficiently high, ensuring enhanced fluorescence.

The degree of circular polarization reaches its maximum where the ratio of optical chirality to electric energy density enhancement is largest in Figures~\ref{reciprocity}c and~\ref{reciprocity}d, consistent with Equation~\ref{DOCP_SymPoint_Cenh_E_enh}. Notably, the dielectric ring significantly boosts the magnetic field enhancement in disk--ring A (up to 35 times) compared to the holey disk (10 times). Because the electric energy enhancement is comparable in both systems at their respective magnetic resonances, the larger magnetic response translates to an approximately 3.5-fold higher polarization in disk--ring A.

The reciprocity-based analytical model accurately predicts the polarization enhancement observed in the numerical simulations. We validate it by comparing the DOCP predicted from the reciprocity-based expression in Equation~\ref{DOCP_SymPoint_Cenh_E_enh} with the results obtained from direct numerical calculations (black and red curves in Figures~\ref{reciprocity}c and~\ref{reciprocity}d). The model further predicts the explicit dependence of the polarization enhancement on the intrinsic dissymmetry parameter $\alpha$ when the magnetic energy contribution is taken into account (Equation~\ref{DOCP_SymPoint}). To illustrate this effect, we plot the DOCP enhancement as a function of $\alpha$ for the holey disk and disk--ring A configurations at their respective wavelengths of maximum polarization (Figure~\ref{reciprocity}e). As $\alpha$ increases, the magnetic and electric energy contributions become comparable, reducing the DOCP enhancement and eventually leading to suppressed polarization as $\alpha$ approaches unity. In contrast, for weakly chiral emitters with small $\alpha$, the magnetic contribution remains negligible and the DOCP enhancement approaches saturation.

The reciprocity framework also reveals how the chiral emitter position affects the measured polarization, exposing intrinsic and extrinsic polarization contributions.
Using Equation~\ref{DOCP_General}, we compute the DOCP at $\lambda_{\mathrm{em}}=560$~nm for different positions along the central horizontal plane of the disk--ring A structure. Each point in Figure~\ref{reciprocity}f (left) represents the DOCP in the normal direction for a right-handed chiral emitter with dissymmetry parameter $\alpha=0.005$ placed at that location.
At the symmetry point (i.e., the center of the hole), the polarization agrees with Equation~\ref{DOCP_SymPoint} (Figures~\ref{reciprocity}c and \ref{reciprocity}d). In contrast, away from the central hole, the DOCP increases significantly, but it is dominated by the nanostructure itself rather than by the intrinsic handedness of the molecular emitter.
To confirm this interpretation, we repeat the analysis for a left-handed chiral emitter with $\alpha=-0.005$ (Figure~\ref{reciprocity}f, right). While the DOCP around the symmetry point reverses sign, the enhanced polarization away from the center remains nearly unchanged for both enantiomers, demonstrating that it is predominantly an extrinsic effect induced by the nanostructure. Similar extrinsic polarization can also emerge when achiral emitters are coupled to other nanostructures~\cite{pfaffenberger2023far}. Positioning the chiral emitter close to the symmetry point of an achiral nanostructure is therefore essential for reliable chiral sensing.

\begin{figure}[t!]
    \centering
     \includegraphics[width=\textwidth]{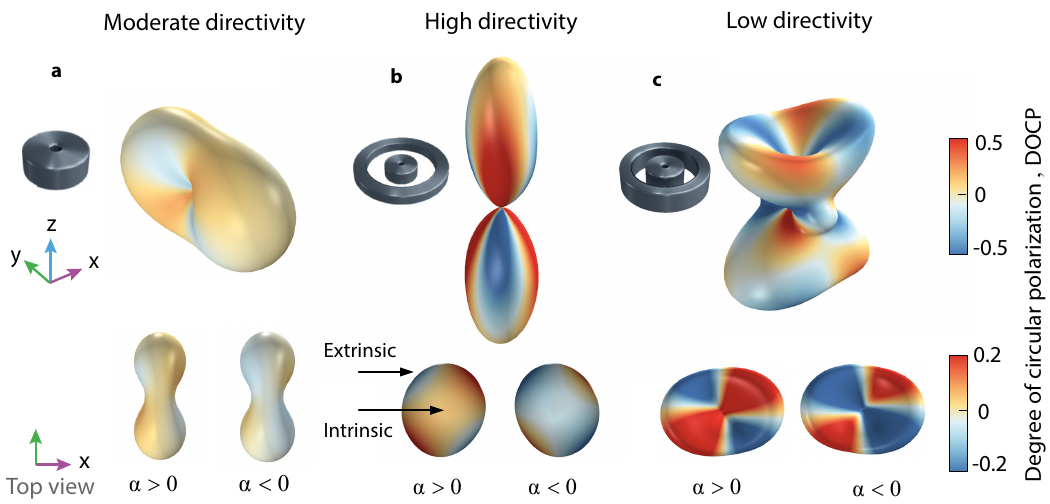}
    \caption{\textbf{Angular patterns with intrinsic and extrinsic polarization}. Radiation patterns with color-coded degree of circular polarization for a chiral emitter with dissymmetry parameter $\alpha = 0.005$, placed at the symmetry point of the three studied configurations: (a) holey disk, and (b,c) disk--ring systems A and B. Their geometrical parameters are as in Figure~\ref{Sf_Exci_Ce_Qe}. Bottom: top views of the radiation patterns for opposite molecular handednesses ($\alpha = \pm0.005$) highlighting intrinsic polarization features that reverse with molecular chirality and extrinsic polarization induced by the nanostructure.}
    \label{Rad_Patterns}
\end{figure}

\section{Angular and multipolar origin of the chiral emission}

Angular radiation patterns also capture the trade-offs between polarization enhancement and collection efficiency.
We present the normalized radiation patterns (far-field intensity) in Figure~\ref{Rad_Patterns} of the three structures introduced in Figure~\ref{Sf_Exci_Ce_Qe} evaluated at their wavelengths of maximum polarization. In all cases, the chiral emitter is placed at the center of the system. The overlaid colormap indicates the degree of circular polarization, and top-view projections below each panel are included to emphasize polarization along the normal direction. The patterns corroborate the trends observed for the three systems in Figure~\ref{Sf_Exci_Ce_Qe}. The highly directive emission of disk--ring A (Figure~\ref{Rad_Patterns}b) leads to enhanced collection efficiency compared to a dipole in free space. By contrast, disk--ring B (Figure~\ref{Rad_Patterns}c) exhibits a larger DOCP of 0.4 in the normal direction, but redirects a significant fraction of the emission to off-normal angles, reducing the collection efficiency.

To distinguish intrinsic polarization arising from molecular chirality from extrinsic polarization induced by the nanostructure, we evaluate the radiation patterns for opposite enantiomers with $\alpha = \pm0.005$ coupled to the resonators. In all cases, the polarization in the normal direction reverses with molecular handedness, which is a direct consequence of positioning the emitter at the symmetry point. A significant polarization is observed at off-normal angles, but this extrinsic contribution arises from the nanostructure rather than from the handedness of the chiral emitter, as it is not reversed for opposite enantiomers. For a dipole located at a symmetry point, this off-normal polarization is related through reciprocity to the background polarization in the normal direction for off-axis dipole positions (see Figure~\ref{reciprocity}f).

\begin{figure} [t!]
    \centering
    \includegraphics[width=\textwidth]{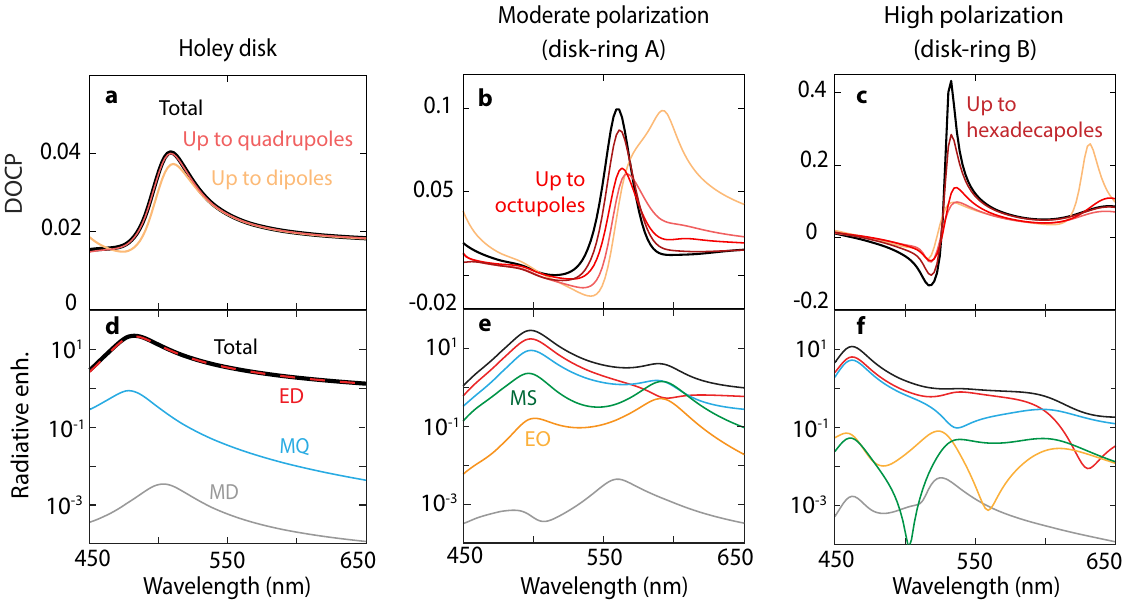}
    \caption{\textbf{Multipolar expansion of circular polarization enhancement.} Multipole contributions to the DOCP emitted in the normal direction and the total radiated power for the three structures in Figure \ref{Sf_Exci_Ce_Qe} excited by a chiral dipole with $\alpha = 0.005$. E and M denote electric and magnetic multipoles. D, Q, O, and S stand for the dipole, quadrupole, octupole, and hexadecapole, respectively. For polarization, the cumulative sum of multipolar fields is considered to identify relevant multipole combinations creating chiral emission. For radiated power, the contributions of individual multipoles are presented to determine the dominant term for emitted intensity.}
    \label{MultD}
\end{figure}

We perform a multipolar decomposition of the radiated fields to identify the contributions of different multipoles to the emitted polarization and intensity (Figure~\ref{MultD}, see Supporting Section~S4 for details)~\cite{muhlig2011multipole, bohren2008absorption}.
We isolate the contributions from the different orders of electric ($E_n$) and magnetic ($M_n$) multipoles to both the DOCP (Figures~\ref{MultD}a--c) and the enhancement of total radiated power (Figures~\ref{MultD}d--f).
Although the radiated power gets separated into distinct multipolar components due to orthogonality, the DOCP can have contributions from cross terms between electric and magnetic multipoles of different orders.
We hence show the cumulative contribution to DOCP from the first $n$ multipoles $(E_{1\rightarrow n}, M_{1\rightarrow n})$.

The multipolar analysis reveals that circular polarization and fluorescence enhancements originate from different mode combinations. For the holey disk operating close to the Kerker condition (Figure ~\ref{reciprocity}a), the polarization arises mainly from the combination of electric and magnetic dipoles (Figure \ref{MultD}a). The radiated power, however, is dominated by the electric dipole (red line in Figure \ref{MultD}d), leading to a doughnut-shaped radiation pattern with low collection efficiency (Figure \ref{Sf_Exci_Ce_Qe}d and Figure \ref{Rad_Patterns}a). Introducing the dielectric ring in the disk--ring systems gives rise to higher-order multipoles that affect both the polarization (Figures~\ref{MultD}b and~\ref{MultD}c) and the radiation pattern (Figures~\ref{MultD}e and~\ref{MultD}f). Multipoles up to hexadecapole order are required to qualitatively reproduce the observed polarization enhancement in these systems. For disk--ring A, the presence of a large magnetic quadrupole accompanied by an electric dipole produces directional radiation at the emission wavelength (Figure \ref{Rad_Patterns}). 

\section{Conclusion}

To push chiral sensitivity down to single molecules, we have introduced a fluorescence-based approach to simultaneously enhance both fluorescence intensity and its polarization. We developed a reciprocity framework describing nanophotonic enhancement of circularly polarized fluorescence for a chiral emitter and a general nanoantenna. Our closed-form expressions link the degree of circular polarization to the local optical chirality and electric energy density, identifying the trade-offs between fluorescence and polarization enhancement.
Building on this analysis, we proposed high-index dielectric nanostructures that combine strong electric and magnetic resonances to enhance the degree of circular polarization and the fluorescence simultaneously.
Their electric resonances increase the excitation rate, quantum efficiency, and collection efficiency, leading to large  fluorescence enhancements. The magnetic resonances increase the optical chirality density and thereby yield high polarization enhancement producing superchiral emission. The proposed nanostructures enable 1000- and 10-fold enhancements of fluorescence intensity and polarization.

Beyond improving the signal-to-noise ratio to increase sensitivity, our analysis clarifies how to reliably identify molecular chirality in nanophotonic environments. The reciprocity formalism and angular emission patterns reveal the coexistence of intrinsic polarization, which reverses with molecular handedness, and extrinsic polarization induced by the nanostructure. Positioning the emitter close to a symmetry point of an achiral resonator enables unambiguous enantiomer discrimination. Measuring in the normal emission direction shows robust polarization avoiding highly polarized angular features of extrinsic origin.
Finally, a multipolar decomposition shows that fluorescence and polarization enhancements originate from different mode combinations. While radiated intensity is often dominated by a small number of individual multipoles, circular polarization emerges from cross terms creating electric--magnetic interference involving multiple multipolar orders.

More broadly, the reciprocity framework, the identification of intrinsic and extrinsic polarization contributions, and the multipolar insights introduced here apply to a wider range of nanophotonic platforms for circularly polarized emission and chiral sensing. 
By engineering brighter and more polarized chiral emission, our results provide a realistic path toward single-molecule chiroptical detection to shed light on biomolecular structure and conformational changes.


\begin{acknowledgement}

This work was financially supported by the European Research Council (ERC) under the European Union's Horizon 2020 Research and Innovation Program (Grant Agreement 948804, CHANSON). We thank Tom T. C. Sistermans for helpful discussions that improved the manuscript.

\end{acknowledgement}

\clearpage
\bibliography{ref.bib}

@book{francotte2006chirality,
  title={Chirality in drug research},
  author={Francotte, Eric and Lindner, Wolfgang},
  volume={33},
  year={2006},
  publisher={Wiley-VCH Weinheim}
}

@article{zhou2018chiral,
  title={Chiral pharmaceuticals: Environment sources, potential human health impacts, remediation technologies and future perspective},
  author={Zhou, Yaoyu and Wu, Shikang and Zhou, Hao and Huang, Hongli and Zhao, Jia and Deng, Yaocheng and Wang, Hua and Yang, Yuan and Yang, Jian and Luo, Lin},
  journal={Environ. Int.},
  volume={121},
  pages={523--537},
  year={2018},
  publisher={Elsevier}
}

@book{kurihara1998chirality,
  title={Chirality in agrochemicals},
  author={Kurihara, Norio and Miyamoto, Junshi},
  year={1998},
  publisher={John Wiley and Sons}
}

@book{crossley1995chirality,
  title={Chirality and biological activity of drugs},
  author={Crossley, Roger J},
  volume={7},
  year={1995},
  publisher={CRC Press}
}

@article{schellman1975circular,
  title={Circular dichroism and optical rotation},
  author={Schellman, John A},
  journal={Chem. Rev.},
  volume={75},
  number={3},
  pages={323--331},
  year={1975},
  publisher={ACS Publications}
}

@book{schaferling2017chiral,
  title={Chiral nanophotonics},
  author={Sch{\"a}ferling, Martin},
  series={Springer Series in Optical Sciences},
  volume={205},
  year={2017},
  publisher={Springer}
}

@article{solomon2020nanophotonic,
  title={Nanophotonic platforms for chiral sensing and separation},
  author={Solomon, Michelle L and Saleh, Amr AE and Poulikakos, Lisa V and Abendroth, John M and Tadesse, Loza F and Dionne, Jennifer A},
  journal={Acc. Chem. Res.},
  volume={53},
  number={3},
  pages={588--598},
  year={2020},
  publisher={ACS Publications}
}

@article{im2024perspectives,
  title={Perspectives of chiral nanophotonics: from mechanisms to biomedical applications},
  author={Im, Seongmin and Mousavi, Seyedehniousha and Chen, Yun-Sheng and Zhao, Yang},
  journal={npj Nanophotonics},
  volume={1},
  pages={46},
  year={2024},
  publisher={Nature Publishing Group UK London}
}

@article{yoo2019metamaterials,
  title={Metamaterials and chiral sensing: a review of fundamentals and applications},
  author={Yoo, SeokJae and Park, Q-Han},
  journal={Nanophotonics},
  volume={8},
  number={2},
  pages={249--261},
  year={2019},
  publisher={De Gruyter}
}

@article{tang2009limits,
  title={Limits on fluorescence detected circular dichroism of single helicene molecules},
  author={Tang, Yiqiao and Cook, Timothy A and Cohen, Adam E},
  journal={J. Phys. Chem. A},
  volume={113},
  number={22},
  pages={6213--6216},
  year={2009},
  publisher={ACS Publications}
}

@article{turner1974fluorescence,
  title={Fluorescence detected circular dichroism},
  author={Turner, Douglas H and Tinoco Jr, Ignacio and Maestre, Marcos},
  journal={J. Am. Chem. Soc.},
  volume={96},
  number={13},
  pages={4340--4342},
  year={1974},
  publisher={ACS Publications}
}

@article{tinoco1977fluorescence,
  title={Fluorescence detected circular dichroism and circular polarization of luminescence in rigid media: Direction dependent optical activity obtained by photoselection},
  author={Tinoco Jr, Ignacio and Ehrenberg, B and Steinberg, IZ},
  journal={J. Chem. Phys.},
  volume={66},
  number={3},
  pages={916--920},
  year={1977},
  publisher={American Institute of Physics}
}

@article{kuhn2006enhancement,
  title={Enhancement of single-molecule fluorescence using a gold nanoparticle as an optical nanoantenna},
  author={K{\"u}hn, Sergei and H{\aa}kanson, Ulf and Rogobete, Lavinia and Sandoghdar, Vahid},
  journal={Phys. Rev. Lett.},
  volume={97},
  number={1},
  pages={017402},
  year={2006},
  publisher={APS}
}

@article{curto2010unidirectional,
  title={Unidirectional emission of a quantum dot coupled to a nanoantenna},
  author={Curto, Alberto G. and Volpe, Giorgio and Taminiau, Tim H. and Kreuzer, Mark P. and Quidant, Romain and van Hulst, Niek F.},
  journal={Science},
  volume={329},
  number={5994},
  pages={930--933},
  year={2010},
  publisher={American Association for the Advancement of Science}
}

@article{tam2007plasmonic,
  title={Plasmonic enhancement of molecular fluorescence},
  author={Tam, Felicia and Goodrich, Glenn P and Johnson, Bruce R and Halas, Naomi J},
  journal={Nano Lett.},
  volume={7},
  number={2},
  pages={496--501},
  year={2007},
  publisher={ACS Publications}
}

@article{cihan2018silicon,
  title={Silicon {M}ie resonators for highly directional light emission from monolayer {MoS$_2$}},
  author={Cihan, Ahmet Fatih and Curto, Alberto G and Raza, S{\o}ren and Kik, Pieter G and Brongersma, Mark L},
  journal={Nat. Photonics},
  volume={12},
  number={5},
  pages={284--290},
  year={2018},
  publisher={Nature Publishing Group}
}

@article{murai2020enhanced,
  title={Enhanced light emission by magnetic and electric resonances in dielectric metasurfaces},
  author={Murai, Shunsuke and Castellanos, Gabriel W. and Raziman, T. V. and Curto, Alberto G. and Rivas, Jaime G{\'o}mez-Rivas},
  journal={Adv. Opt. Mater.},
  pages={1902024},
  volume={8},
  number={16},
  year={2020},
  publisher={Wiley Online Library}
}

@article{raziman2019conditions,
  title={Conditions for enhancing chiral nanophotonics near achiral nanoparticles},
  author={Raziman, T V and Godiksen, Rasmus H and M{\"u}ller, Moos A and Curto, Alberto G},
  journal={ACS Photonics},
  volume={6},
  number={10},
  pages={2583--2589},
  year={2019},
  publisher={ACS Publications}
}

@article{feng2017isotropic,
  title={Isotropic magnetic Purcell effect},
  author={Feng, Tianhua and Zhang, Wei and Liang, Zixian and Xu, Yi and Miroshnichenko, Andrey E},
  journal={ACS Photonics},
  volume={5},
  number={3},
  pages={678--683},
  year={2017},
  publisher={ACS Publications}
}

@article{bidault2019dielectric,
  title={Dielectric nanoantennas to manipulate solid-state light emission},
  author={Bidault, S{\'e}bastien and Mivelle, Mathieu and Bonod, Nicolas},
  journal={J. Appl. Phys.},
  volume={126},
  number={9},
  pages={094104},
  year={2019},
  publisher={AIP Publishing LLC}
}

@article{yang2017boosting,
  title={Boosting magnetic field enhancement with radiative couplings of magnetic modes in dielectric nanostructures},
  author={Yang, Zhong-Jian and Zhao, Qian and He, Jun},
  journal={Opt. Express},
  volume={25},
  number={14},
  pages={15927--15937},
  year={2017},
  publisher={Optical Society of America}
}

@article{zambrana2016tailoring,
  title={Tailoring the chirality of light emission with spherical Si-based antennas},
  author={Zambrana-Puyalto, Xavier and Bonod, Nicolas},
  journal={Nanoscale},
  volume={8},
  number={19},
  pages={10441--10452},
  year={2016},
  publisher={Royal Society of Chemistry}
}

@article{tang2011enhanced,
  title={Enhanced enantioselectivity in excitation of chiral molecules by superchiral light},
  author={Tang, Yiqiao and Cohen, Adam E},
  journal={Science},
  volume={332},
  number={6027},
  pages={333--336},
  year={2011},
  publisher={American Association for the Advancement of Science}
}

@book{balanis1999advanced,
  title={Advanced Engineering Electromagnetics},
  author={Balanis, Constantine A},
  year={1999},
  publisher={John Wiley \& Sons}
}

@article{tang2010optical,
  title={Optical chirality and its interaction with matter},
  author={Tang, Yiqiao and Cohen, Adam E},
  journal={Phys. Rev. Lett.},
  volume={104},
  number={16},
  pages={163901},
  year={2010},
  publisher={APS}
}

@article{sun2016fluorescence,
  title={Fluorescence enhancement in visible light: dielectric or noble metal?},
  author={Sun, S and Wu, L and Bai, P and Png, CE},
  journal={Phys. Chem. Chem. Phys.},
  volume={18},
  number={28},
  pages={19324--19335},
  year={2016},
  publisher={Royal Society of Chemistry}
}

@article{mohammadi2008gold,
  title={Gold nanorods and nanospheroids for enhancing spontaneous emission},
  author={Mohammadi, Ahmad and Sandoghdar, Vahid and Agio, Mario},
  journal={New J. Phys.},
  volume={10},
  number={10},
  pages={105015},
  year={2008},
  publisher={IOP Publishing}
}

@article{mohammadi2019accessible,
  title={Accessible superchiral near-fields driven by tailored electric and magnetic resonances in all-dielectric nanostructures},
  author={Mohammadi, Ershad and Tavakoli, Ahad and Dehkhoda, Parisa and Jahani, Yasaman and Tsakmakidis, Kosmas L and Tittl, Andreas and Altug, Hatice},
  journal={ACS Photonics},
  volume={6},
  number={8},
  pages={1939--1946},
  year={2019},
  publisher={ACS Publications}
}

@article{mohammadi2018nanophotonic,
  title={Nanophotonic platforms for enhanced chiral sensing},
  author={Mohammadi, E and Tsakmakidis, KL and Askarpour, AN and Dehkhoda, P and Tavakoli, A and Altug, H},
  journal={ACS Photonics},
  volume={5},
  number={7},
  pages={2669--2675},
  year={2018},
  publisher={ACS Publications}
}

@article{aspnes1983dielectric,
  title={Dielectric functions and optical parameters of {Si}, {Ge}, {GaP}, {GaAs}, {GaSb}, {InP}, {InAs}, and {InSb} from 1.5 to 6.0 {eV}},
  author={Aspnes, D E and Studna, A A},
  journal={Phys. Rev. B},
  volume={27},
  number={2},
  pages={985},
  year={1983},
  publisher={APS}
}

@Article{wientjes2014enhance,
author ="Wientjes, Emilie and Renger, Jan and Curto, Alberto G. and Cogdell, Richard and van Hulst, Niek F.",
title  ="Nanoantenna enhanced emission of light-harvesting complex 2: the role of resonance{,} polarization{,} and radiative and non-radiative rates",
journal  ={Phys. Chem. Chem. Phys.},
year  ="2014",
volume  ="16",
issue  ="45",
pages  ="24739-24746",
publisher  ="The Royal Society of Chemistry",
doi  ="10.1039/C4CP03636K",
}

@article{gaiduk2010room,
  title={Room-temperature detection of a single molecule’s absorption by photothermal contrast},
  author={Gaiduk, A and Yorulmaz, M and Ruijgrok, PV and Orrit, M},
  journal={Science},
  volume={330},
  number={6002},
  pages={353--356},
  year={2010},
  publisher={American Association for the Advancement of Science}
}

@article{muhlig2011multipole,
  title={Multipole analysis of meta-atoms},
  author={M{\"u}hlig, Stefan and Menzel, Christoph and Rockstuhl, Carsten and Lederer, Falk},
  journal={Metamaterials},
  volume={5},
  number={2-3},
  pages={64--73},
  year={2011},
  publisher={Elsevier}
}

@book{bohren2008absorption,
  title={Absorption and scattering of light by small particles},
  author={Bohren, Craig F and Huffman, Donald R},
  year={2008},
  publisher={John Wiley \& Sons}
}

@article{anger2006fluor,
  title = {Enhancement and Quenching of Single-Molecule Fluorescence},
  author = {Anger, Pascal and Bharadwaj, Palash and Novotny, Lukas},
  journal = {Phys. Rev. Lett.},
  volume = {96},
  issue = {11},
  pages = {113002},
  numpages = {4},
  year = {2006},
  month = {Mar},
  publisher = {American Physical Society},
  doi = {10.1103/PhysRevLett.96.113002}
}

@article{solomon2020fluorescence,
  title={Fluorescence-detected circular dichroism of a chiral molecular monolayer with dielectric metasurfaces},
  author={Solomon, Michelle L and Abendroth, John M and Poulikakos, Lisa V and Hu, Jack and Dionne, Jennifer A},
  journal={J. Am. Chem. Soc.},
  volume={142},
  number={43},
  pages={18304--18309},
  year={2020},
  publisher={ACS Publications}
}

@article{mohammadi2021dual,
author = {Mohammadi, Ershad and Tittl, Andreas and Tsakmakidis, Kosmas L. and Raziman, T. V. and Curto, Alberto G.},
title = {Dual nanoresonators for ultrasensitive chiral detection},
journal = {ACS Photonics},
volume = {8},
number = {6},
pages = {1754-1762},
year = {2021},
doi = {10.1021/acsphotonics.1c00311},
}

@article{graf2019achiral,
  title={Achiral, helicity preserving, and resonant structures for enhanced sensing of chiral molecules},
  author={Graf, Florian and Feis, Joshua and Garcia-Santiago, Xavier and Wegener, Martin and Rockstuhl, Carsten and Fernandez-Corbaton, Ivan},
  journal={ACS Photonics},
  volume={6},
  number={2},
  pages={482--491},
  year={2019},
  publisher={ACS Publications}
}

@article{eismann2018exciting,
  title={Exciting a chiral dipole moment in an achiral nanostructure},
  author={Eismann, J{\"o}rg S and Neugebauer, Martin and Banzer, Peter},
  journal={Optica},
  volume={5},
  number={8},
  pages={954--959},
  year={2018},
  publisher={Optical Society of America}
}

@article{darvishzadeh2019optical,
  title={Optical magnetic field enhancement at nanoscale: a nanoantenna comparative study},
  author={Darvishzadeh-Varcheie, Mahsa and Kamandi, Mohammad and Albooyeh, Mohammad and Capolino, Filippo},
  journal={Opt. Lett.},
  volume={44},
  number={20},
  pages={4957--4960},
  year={2019},
  publisher={Optica Publishing Group}
}

@article{li2017enhancing,
  title={Enhancing magnetic dipole emission by a nano-doughnut-shaped silicon disk},
  author={Li, Jiaqi and Verellen, Niels and Van Dorpe, Pol},
  journal={ACS Photonics},
  volume={4},
  number={8},
  pages={1893--1898},
  year={2017},
  publisher={ACS Publications}
}

@article{yang2020strong,
  title={Strong magnetic field enhancement and magnetic Purcell effect in a dielectric disk-ring composite nanocavity},
  author={Yang, Yang and Zhu, Bofeng and Dai, Haitao},
  journal={J. Opt. Soc. Am. B},
  volume={37},
  number={3},
  pages={702--708},
  year={2020},
  publisher={Optica Publishing Group}
}

@article{pfaffenberger2023far,
  title={Far-Field Polarization Optics Control the Nanometer-Scale Pattern of High-Fluorescence Dissymmetry Emission from Achiral Molecules near Plasmonic Nanodimers},
  author={Pfaffenberger, Zechariah J and Chattopadhyay, Saaj and Biteen, Julie S},
  journal={J. Phys. Chem. C},
  year={2023},
  publisher={ACS Publications}
}

@article{arrico2021quantifying,
  title={Quantifying the overall efficiency of circularly polarized emitters},
  author={Arrico, Lorenzo and Di Bari, Lorenzo and Zinna, Francesco},
  journal={Chem. Eur. J.},
  volume={27},
  number={9},
  pages={2920--2934},
  year={2021},
  publisher={Wiley Online Library}
}

@article{mohammadi2023nanophotonic,
  title={Nanophotonic chirality transfer to dielectric Mie resonators},
  author={Mohammadi, Ershad and Raziman, TV and Curto, Alberto G},
  journal={Nano Lett.},
  volume={23},
  number={9},
  pages={3978--3984},
  year={2023},
  publisher={ACS Publications}
}

\end{document}